\documentclass[runningheads]{llncs}
\usepackage{graphicx}
\usepackage{makecell}
\usepackage{dirtytalk}
\usepackage{supertabular}
\usepackage{enumitem}
\usepackage{booktabs}
\usepackage{rotating}
\usepackage{multirow}
\usepackage{tabularx}
\usepackage{amsmath}
\usepackage{ulem}
\usepackage{caption}
\usepackage{subcaption}
\usepackage{array}
\newcolumntype{P}[1]{>{\centering\arraybackslash}p{#1}}
\newcolumntype{M}[1]{>{\centering\arraybackslash}m{#1}}

\begin{document}
\title{Quantifying Residual Motion Artifacts \\in Fetal fMRI Data}

\author{Athena Taymourtash\inst{1} \and
Ernst Schwartz\inst{1} \and Karl-Heinz Nenning\inst{1} \and Daniel Sobotka\inst{1}\and Mariana Diogo\inst{2}\and Gregor Kasprian\inst{2} \and Daniela Prayer \inst{2} \and Georg Langs\inst{1}\thanks{This project has received funding from the European Union's Horizon 2020 research and innovation programme under the Marie Sk\l odowska-Curie grant agreement No 765148.}} 
\authorrunning{A. Taymourtash et al.}
\institute{Computational Imaging Research Lab, Department of Biomedical Imaging and Image-guided Therapy, Medical University of Vienna, Vienna, Austria 
\and
Division for Neuroradiology and Musculoskeletal Radiology, Department of Biomedical Imaging and Image-guided Therapy, Medical University of Vienna,
Vienna, Austria
\email{athena.taymourtash@meduniwien.ac.at}\\
}
\maketitle              
\begin{abstract}
\vspace{-5mm}
Fetal functional Magnetic Resonance Imaging (fMRI) has emerged as a powerful tool for investigating brain development in utero, holding promise for generating developmental disease biomarkers and supporting prenatal diagnosis. However, to date its clinical applications have been limited by unpredictable fetal and maternal motion during image acquisition. Even after spatial realigment, these cause spurious signal fluctuations confounding measures of functional connectivity and biasing statistical inference of relationships between connectivity and individual differences. As there is no ground truth for the brain's functional structure, especially before birth, quantifying the quality of motion correction is challenging. In this paper, we propose evaluating the efficacy of different regression based methods for removing motion artifacts after realignment by assessing the residual relationship of functional connectivity with estimated motion, and with the distance between areas. Results demonstrate the sensitivity of our evaluation's criteria to reveal the relative strengths and weaknesses among different artifact removal methods, and underscore the need for greater care when dealing with fetal motion.

\end{abstract}

\section{Introduction}
For over two decades, it has been known that motion artifacts cause serious disruptions to fMRI data such that even sub-millimeter head movement can add spurious variance to true signal and bias inter-individual differences in fMRI metrics \cite{power2012spurious,van2012influence}. The severity of this problem is especially pronounced in fetal imaging due to unpredictable fetal movement, maternal respiration, and signal non-uniformities \cite{malamateniou2013motion}. To date, advanced motion correction methods, often relying on super resolution techniques, have been proposed to address motion artifact by reconstructing a high resolution motion-free volume from several clinical low resolution MR images of a moving fetus \cite{kuklisova2012reconstruction,ebner2018automated}. As methods to counteract motion artifacts are being developed, it is of critical importance to know if a technique has improved the quality of data or introduced additional artifacts. Among several quality-control benchmarks that have been recently employed in adult studies \cite{ciric2017benchmarking,parkes2018evaluation,lydon2019evaluation}, Quality Control-Functional Connectivity (QC-FC) correlation was found to be the most useful metric of quality as it directly quantifies the relationship between motion and the primary outcome of interest over a population~\cite{power2012spurious}. The QC-FC benchmark is based on the correlation between the FC of each pair of regions and the average motion of each subjects in the dataset to determine how that connectivity is modulated by subject motion. Since both FC and motion are calculated as a mean value over the entire scan, for the rest of paper we call it \textit{static FC-FD}. For the purpose of assessing residual artifacts in fetuses, the average motion is insufficient, due to excessive fetal motion, exhibiting large movement spikes and overall more continuous motion during acquisition \cite{malamateniou2013motion}. Furthermore, using the average motion as a means to quantify spurious connectivity allows for no subject-specific evaluation as it provides only group specific motion dependencies. It therefore is not able to decide whether a specific acquisition should be removed from analysis entirely, or could be salvaged by excluding specific contaminated time-points using methods such as scrubbing \cite{power2014methods}. 

In this paper, we developed a \textit{dynamic FC-FD} benchmark for systematic evaluation of subject-specific fMRI data quality, comparing the efficacy of existing regression strategies for mitigating motion-induced artefacts. We evaluated our benchmark on fetal fMRI as an application with irregular motion. However, the proposed methodology is general and can be applied to any fMRI study.

\section{Data, preprocessing and motion correction} 
Experiments in this study were performed on 24 in-utero BOLD MRI sequences obtained from fetuses between 19 and 39 weeks of gestation. None of the cases showed any neurological pathologies. Pregnant women were scanned on a 1.5T clinical scanner (Philips Medical Systems, Best, Netherlands) using single-shot echo-planar imaging (EPI), and a sensitivity encoding (SENSE) cardiac coil with five elements. Image matrix size was 144$\times$144, with 3$\times$3$mm^{2}$ in-plane resolution, 3$mm$ slice thickness, a TR/TE of 1000/50 ms, and a flip angle of 90. Each scan contains 96 volumetric images obtained in an interleaved slice order to minimize cross-talk between adjacent slices. 
Preprocessing of the resting-state data included correction for distortions induced by magnetic field inhomogeneity, slice timing correction, motion correction, de-meaning and removal of any linear or quadratic trends. Motion correction comprised iterative rigid-body registration of all slice stacks to a resulting mean volume so that the objective function of normalized correlation ratio was optimized. After 25 iterations, a realigned version of each volume was created using trilinear interpolation, and two slices were interpolated between every two slices to eliminate the effect of slice interleaving in different stacks, tripling the slice number after motion correction.    

\section{Functional connectivity after nuisance regression}

Individual functional connectivity analysis was performed in the native functional space. For this, cortical regions of interest (ROIs) were first obtained using an automatic atlas-based segmentation of $T2$ scans of the same subject acquired during the same scan session as the fMRI volumes, using a publicly available atlas of fetal brain anatomy \cite{gholipour2017normative}. The resulting parcellation consists of 98 ROIs and was mapped to the motion corrected fMRI space using a rigid transformation computed between each individual structural T2 scan and the first volume of fMRI data. For each parcel, we calculated the mean time course of all voxels, and applied one of eight different common nuisance regression strategies (Table~\ref{tab1}). The resulting time course served as basis for calculating functional connectivity in the form of a correlation matrix estimated using Pearson's correlation.

\begin{table}[t]
 \caption{Eight nuisance regression strategies evaluated.}\label{tab1}
 
\setlength{\tabcolsep}{8pt}
 
 \begin{small}
 \begin{tabular}{>{\raggedright\arraybackslash}p{2cm}p{8cm}M{.4cm}}
 \toprule
 \textbf{Strategy} & \textbf{Summary of regressors} & \textbf{\#R}\\
 \midrule
\textbf{1:} GSR & {Mean time-series averaged across the entire brain}  & 1 \\  
 \hline
\textbf{2:} 2Phys & Two physiological time-series computed across white matter (WM) and cerebrospinal fluid (CSF). & 2 \\ 
 \hline
\textbf{3:} 6HMP & 6 motion parameter estimates derived from realignment. & 6\\ 
 \hline
\textbf{4:} 6HMP +~2Phys~+~GSR & 6 motion parameter estimates, 2 physiological  compartments and GSR & 9 \\ \hline
\textbf{5:} 24HMP & 6 motion parameters, their temporal derivatives, together with quadratic expansions of parameters and derivatives. & 24 \\ 
 \hline
\textbf{6:} 24HMP + 8Phys~+~4GSR & Quadratic expansion of model 4: 9 regressors, their derivatives, quadratic terms, and squares of derivatives.  & 36 \\ 
 \hline
\textbf{7:} aCompCor & 5 principal components each from the WM and CSF~\cite{behzadi2007component} & 10 \\ 
 \hline
\textbf{8:} tCompCor & 6 principal components from high-variance voxels~\cite{behzadi2007component} & 6 \\ 
  \bottomrule
 \end{tabular}
 \end{small}
 \end{table}

\section{Assessing spurious motion artifacts in fMRI data}

Motion correction first aims to re-align individual slices of the fMRI volume sequence such that the anatomical position of a voxel is consistent for the whole time-series. Then, information such as spatial displacement and other surrogate measurements of non-neural signal are used as nuisance regressors to remove non-physiological residual signal components. We propose a dynamic FC-FD method based on the fact that the sources of motion in fMRI time series are non-stationary and can potentially induce changes in FC over time \cite{lurie2018nature,hindriks2016can}. It expands on subject level \textit{static FC-FD}~\cite{power2012spurious} by taking variation of motion and FC over time into account.
We use the association between FC variation and motion before and after the application of each nuisance regression strategy to evaluate its effectiveness in fMRI data of a subject. 
FC varies over time due to noise and actual non-stationary neural behavior with the magnitude of variation not differing from simulated stochastic time-series \cite{hindriks2016can}. 
Nevertheless, we can exploit the relationship between FC fluctuation, estimated motion, and the distance between areas to quantify possible residual non-neural confounds in the signal after motion correction and nuisance regression. Here we focus on comparing different variants of the latter based on several measures. 
It includes three steps: quantifying FC variation, measuring a subject-specific motion time-course, and evaluating the association between these two together with the distance between regions.

\subsection{Capturing fluctuations of connectivity with a sliding window}

We estimate fluctuating connectivity over time using a sliding window approach, resulting in a vector of FC values for every pair of regions. Ideally, the window should be large enough to permit robust estimation of FC, yet small enough to detect transient effects properly. We extracted 50 overlapping windows for each time series, corresponding to a duraction of 46s and 1s step-size of the sliding window. This is consistent with the majority of previously published values ranging from 30- to 60-s \cite{preti2017dynamic}. Finally, a Fisher z-transformation was applied to all correlations, resulting in a three-dimensional ($98\times98\times50$) tensor of FC values for each subject.

\subsection{Capturing head motion as framewise displacement vector}

For measuring subject-specific motion, we used rigid body realignment estimates obtained from motion correction step. These six realignment parameters (translation: $x$, $y$, $z$; rotation: $\alpha$, $\beta$, $\gamma$) describe the relative displacement of every volume from a fixed reference volume in the scan. Based on these parameters we calculated framewise displacement (FD) as proposed in \cite{jenkinson2002improved}. For each sliding window, we then computed the average FD to quantify a dynamic subject-specific vector of head motion. 

\subsection{Evaluating the association between FC and motion}
Static association is measured across the study population using the correlation of mean functional connectivity and framewise displacement averaged over the whole fMRI scan \cite{power2012spurious}. For example, 24 fetuses in our study would yield 24 mean FD and 24 FC values for a specific edge in their FC map. The correlation between these two is used as surrogate for the modulated of this edge by subject motion. \cite{power2014methods} provides an extended rationale. To take the dynamicity of motion within each fMRI sequence into account, we can calculate FC-FD association analogously on measurements in sliding windows. 
For each pair of regions, we calculated correlation between FC and FD across the sequence of sliding windows, resulting in a \textit{dynamic FC-FD association}.
This captures the subject-specific residual effect of motion for each pair of regions.  To assess the efficacy of nuisance regression strategy, we compared the proportion of edges with significant correlations of FC and FD as well as the median absolute value of their distribution. Fewer significant correlations or lower absolute median of correlations are indicative of better performance. 
\subsection{Evaluating the association between distance and FC-FD correlations}
Previous studies have shown that in-scanner movement primarily inflates short-range FC while decreases long-range connectivity \cite{power2012spurious,van2012influence}. Motion thus affects more severely the FC of short range connections, and the correlation between 
\begin{figure}[ht!]
     \iftrue
     \begin{center}
     \includegraphics[width=1\textwidth]{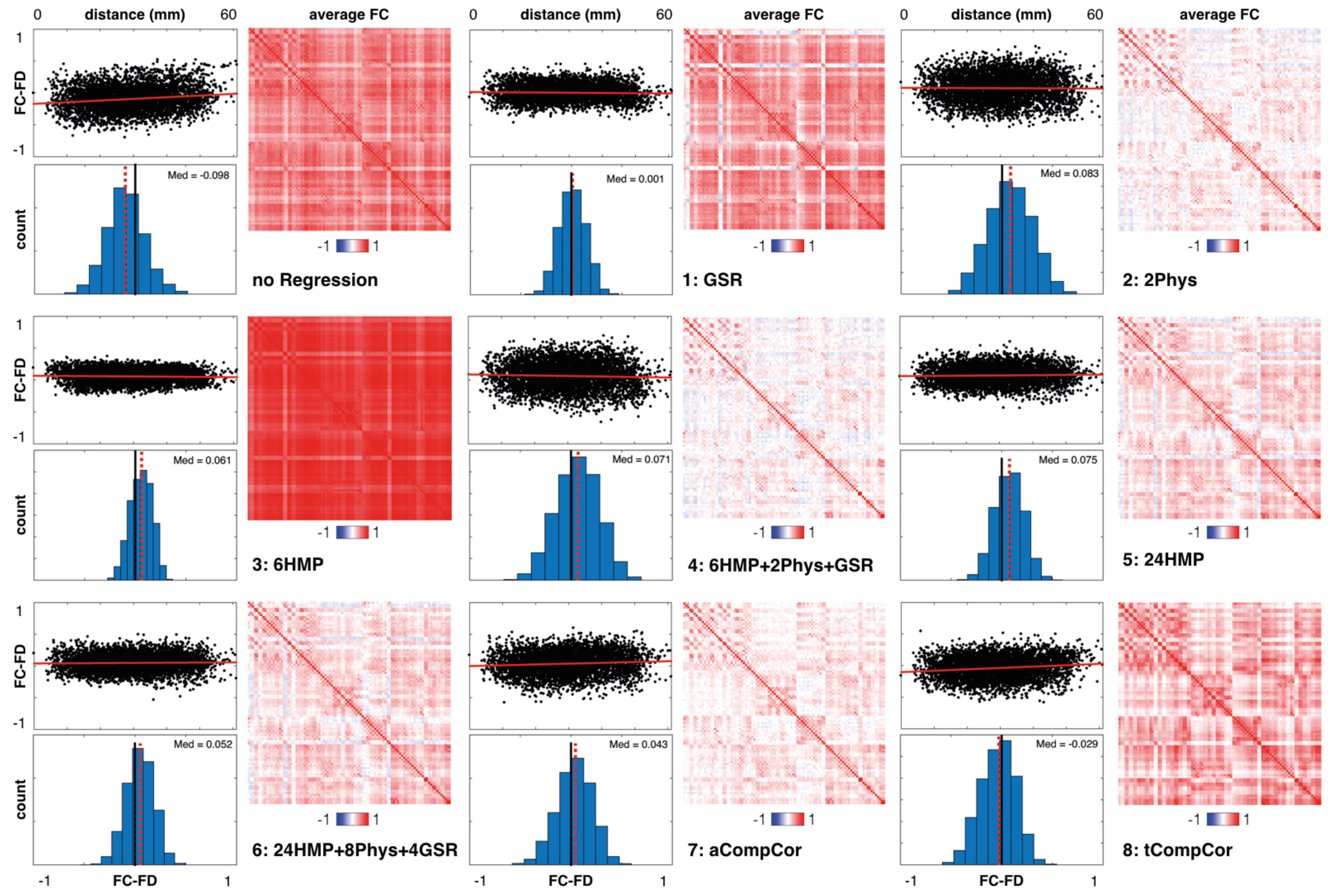}
     \end{center}
     \fi
     \vspace{-3mm}
        \caption{A static benchmark reveals associations between FC and motion (FC-FD), and the association between distance and FC-FD (FC-FD-D) on a population level. Note that every point in the FC-FD-D plots shows the correlation between the FC of one specific edge and the average FD over the whole sample. Quantitative values can be found in Tab.\,\ref{tab2}. According to this benchmark, 6HMP outperforms other strategies, as no significant FC-FD association remained. However, the average FC map after 6HMP exhibits dramatically increased  FC values across the cortex.}
        \label{fig:1}
        
\end{figure}
FC-FD association and distance of region is a possible marker of residual motion artifacts. To determine the residual distance-dependence effects of motion on FC variation, We calculated the distance $D_{ij}$ between regions based on the center of mass of each parcel resulting in distance matrix $D$. We then calculated the correlation between the distance between each pair of parcels and the corresponding FC-FD correlation, we call this \textit{FC-FD-D association}.
\section{Results}
Fetal head motion were quantifed by framewise displacement ranged from $\sim$0 to 43.06 [mm, average: 1.84 $\pm$ 2.12]. To evaluate possible covariation of fetal age and the estimated motion, we measured the correlation between gestational week of subjects and both mean and maximum FD. Although in our study cohort, neither mean FD ($r=0.16$, $p=0.43$) nor maximum FD ($r=0.33$, $p=0.12$) doesn't show significant correlation with gestational age. 

\begin{table}[t]
 \caption{The summary metrics of the static FC-FD/-D association benchmark for each nuisance regression method.}\label{tab2}
 \begin{center}
 \setlength{\tabcolsep}{5pt}
 \begin{tabular}{lrcr}
 \toprule
 Pipeline & \makecell{$\#$Edges related \\ to motion (\%)}  & \makecell{Absolute median \\ FC-FD assoc.}&  \makecell{FC-FD-D \\ assoc.} \\
 \midrule
 No Regression & 242 (5.09) & -0.099 & $0.007$ \\
 1: GSR &  8 (0.17) & 0.001 & $-0.03$ \\ 
  2: 2Phys &  361 (7.60) & 0.083 & $-0.01$\\
  3: 6HMP & 0 (0.00) & 0.061 & $-0.03$\\
  4: 6HMP+2Phys+GSR & 283 (5.95) & 0.071 & $-0.04$\\
  5: 24HMP & 40 (0.84) & 0.075 & $0.029$\\
  6: 24HMP+8Phys+4GSR & 15 (0.31) & 0.052 & 0.02\\
  7: aCompCor & 95 (2.00) & 0.043 & $0.08$\\
  8: tCompCor & 103 (2.16) & 0.029 & $0.14$\\
\bottomrule
 \end{tabular}
 \end{center}
 \end{table}
\subsubsection{Static benchmark of FC-FD/-D associations}  
 The result of static FC-FD analysis is illustrated in Fig.~\ref{fig:1}, where for each method, in the top panel FC-FD correlations for all possible pairs of parcels are plotted against their Euclidean distance. 
The bottom panel indicates the distribution of static FC-FD correlations across the study cohort, and the right panel shows the average connectivity matrix of the study cohort. 
According to this benchmark, all regression techniques were very effective in removing the effects of head motion on FC, reducing the number of connections that were significantly related to motion to less than 8\% with the corresponding absolute median ranging from 0.01 to 0.001. In addition, very small correlation values of the static FC-FD-D association suggest that the relationship between motion, FC, and distance has become negligible after applying regression techniques (see Table~\ref{tab2}). However, the average connectivity matrices are different, and suggest that the resulting FC still carries motion effects, or that these artifacts have even be increased. 
Notably, the regression with realignment parameters (3:6HMP) yielded all FC-FD correlations near zero and removed the distance dependent slope and positive offset in the FC-FD measure, whereas it is obvious from the resulting average FC matrix that FC values are entirely dominated by motion-induced variance, resulting in a strong increase in connectivity among brain regions, regardless of their distance. This suggests that motion parameter estimates were not accurate, or linear regression is not a suitable strategy to remove associated signal components, and in both cases \textit{static FC-FD} couldn't correctly reveal the residual effects of motion on data.

Therefore, although this benchmark has been successfully used in adult studies, it doesn't establish reliable results for fetal studies. The most likely explanation is that in contrast to adult studies where subjects span a wide range of mean FD values, all fetuses show similarly high levels of motion. Hence, the resulting FD vector for adult studies covers a large range of variability, allowing FC-FD correlations to reveal distance-dependence, however, the narrow-ranged FD vector over the fetal sample cannot adequately account for subject's motion and so the reliability of FC-FD correlations would be questionable.
\begin{figure}[tb!]
\begin{center}\includegraphics[width=0.8\textwidth]{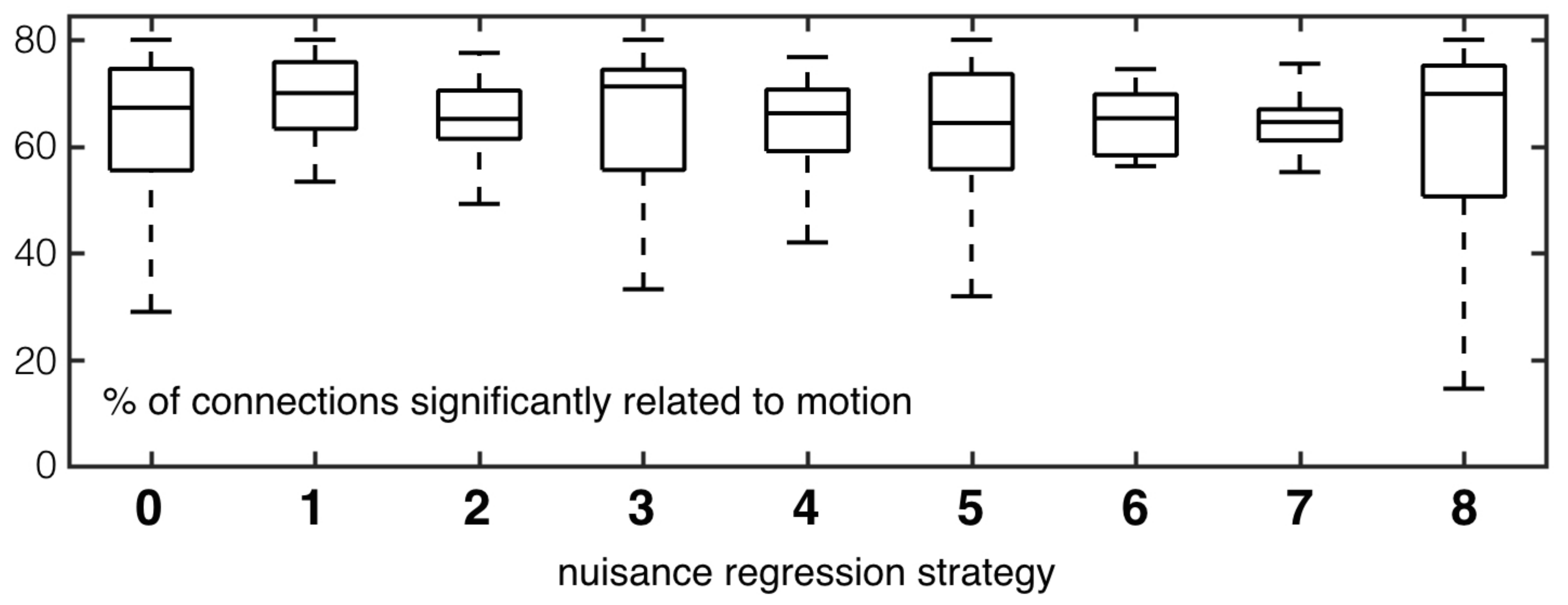}\end{center}

\vspace{-5mm}\caption{Dynamic FC-FD association: the proportion of functional connectivity variations that showed significant associations ($p\leq.05$, uncorrected) with subject-motion after nuisance regression. Fewer significant correlations is indicative of better performance.} \label{fig2}
\end{figure}

\subsubsection{Dynamic FC-FD association}
The correlation between head motion and FC variation was measured for each subject independently across the sliding windows to see if performing a certain regression strategy has decreased the effect of motion on the data or induced additional artificial variance. 
Fig.~\ref{fig2} shows the percentage of network connections where a significant relationship (correlations with $p$-value $\leq$ .05, uncorrected) with motion was present. The benchmark suggests no regression strategy was effective, leaving the majority of network edges with a residual relation with motion. However, aCompCor showed more homogeneous performance over subjects and the commonly used regression strategy relying only on the six motion parameters fared the worst. 
\begin{figure}[ht!]
      \iftrue
     \begin{center}
     \includegraphics[width=1\textwidth]{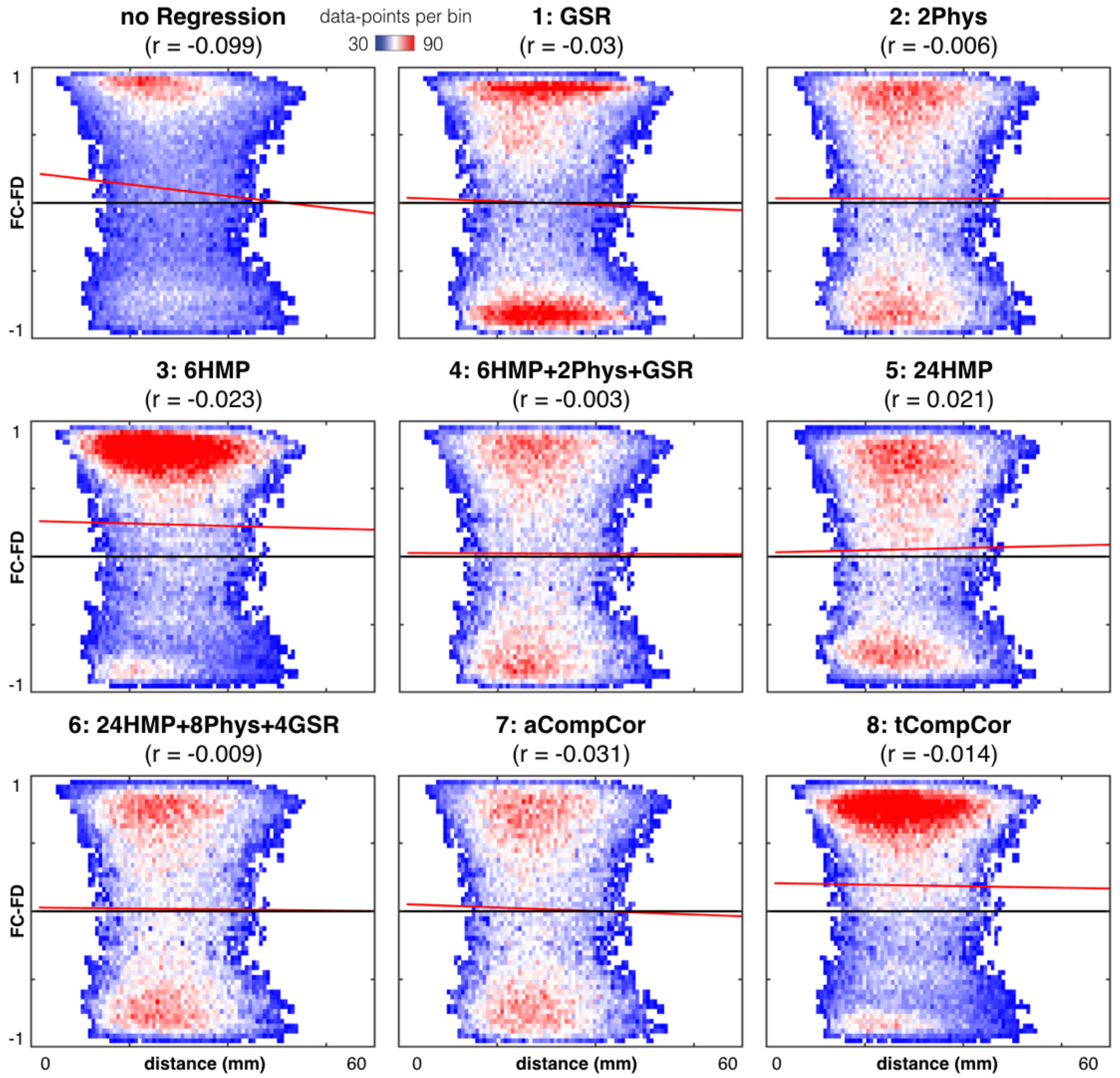}
     \end{center}
     \fi
     \vspace{-3mm}
      \caption{ 
      A density plot of FC-FD association over distance between regions. Trendlines are shown in red. A successful strategy should remove the distance dependent slope and positive offset in of FC-FD vs. distance. Besides, there should be a more density of datapoints around zero axis for an effective strategy. Consistent with previous measure, aCompCor performed better than other strategies, however, this dynamic FC-FD/-D benchmark reveals that clear remaining signs of motion artifacts are still present in datasets.  
      }
        \label{fig3}
\end{figure}
\subsubsection{Dynamic FC-FD-D association}
The benchmark yields an assessment for each subject, with 4753 FC-FD associations per subject. To provide an interpretable visualization of the relationship between edge-wise FC-FD association and region distance, we use a binned scatter plot indicating the density of points with color~(Fig.~\ref{fig3}).

Using no regression model shows that motion influenced BOLD signal in proximal regions homogeneously, resulting in spuriously inflated correlations among those regions. The least effective methods in mitigating such effect were tCompCor and 6HMP. As expected, global signal regression introduced negative correlations due to spurious anti-correlations \cite{murphy2009impact}. aCompCor and the 24HMP+8Phys+4GSR model showed overall better performance relative to other regression strategies. However, the latter is the costliest strategy in terms of the loss of temporal degrees of freedom leading to less statistical confidence in the analysis of fMRI data.
\section{Discussion}
Quality control of fetal fMRI is of utmost importance since its susceptibility to motion artifacts can result in false  observations. A variety of regression strategies provides a choice for removal of non-neural fMRI signal components. As it increasingly becomes more common to use hundreds or even thousands of scans for a single study, it is not practical to manually assess data quality, and in addition, manual assessments are biased and suffer from lack of reproducibility. Group-wise assessment of motion artifacts, such as \textit{static FC-FD}, can be deceptive when there are excessive motion spikes, or generally high motion across the entire population, leading to a ceiling-like effect of motion on correlation values \cite{power2014methods}. Here, we present a dynamic FC-FD/-D benchmark for single-subject fMRI acquisitions that at the same time enables the comparison of nuisance regression approaches. The proposed method was applied to fetal fMRI scans as an application of particularly pronounced and irregular motion. Results suggest that a \textit{static FC-FD} benchmark is not suitable for fetal fMRI studies, as it is not able to capture the relationship between fetal motion and FC, leading to false negative results.

A general limitation of benchmarks evaluating the association with motion is their dependency on realignment based estimates of movement. This is particularly challenging for irregular and substantial motion of fetuses moving inside the uterus and exhibiting spurious large motion spikes. We used one motion correction algorithm to re-align the image data and obtain movement parameter estimates, and better results might be obtained by a different approach. The scope of the paper is the comparison of nuisance regression approaches given likely imperfect movement parameter estimates, and not the comparison of the motion correction approaches themselves. A limiting factor of fetal fMRI is the typically shorter acquisition time, resulting in limitations of FC reliability regardless of motion. Never-the-less quantitative assessment of motion related artifacts is feasible, and dynamic measures as those evaluated in this paper do offer a means to compare methods. In summary, we presented a benchmark for the efficacy of nuisance regression. Results suggest that while they improve the signal, they are not yet adequately effective in removing motion-related variance, given the used motion correction approach.

\bibliographystyle{splncs04}
\bibliography{main.bib}
\end{document}